\documentclass[lettersize,journal]{IEEEtran}
\usepackage{amsmath,amsfonts}
\usepackage{algorithmic}
\usepackage{array}
\usepackage[caption=false,font=normalsize,labelfont=sf,textfont=sf]{subfig}
\usepackage{textcomp}
\usepackage{stfloats}
\usepackage[dvipsnames]{xcolor}
\usepackage{url}
\usepackage{verbatim}
\usepackage{graphicx}
\usepackage{times}
\usepackage{soul}
\usepackage{url}
\usepackage[hidelinks]{hyperref}
\usepackage[utf8]{inputenc}
\usepackage{graphicx}
\usepackage{amsthm}
\usepackage{amssymb}
\usepackage{amsmath}
\usepackage{booktabs}
\usepackage{algorithm}
\usepackage{cite}
\graphicspath{{pics/}}
\usepackage{multirow}
\usepackage{hyperref}
\usepackage[switch]{lineno}
\definecolor{ccr}{RGB}{10, 90,180}
\def\BibTeX{{\rm B\kern-.05em{\sc i\kern-.025em b}\kern-.08em
    T\kern-.1667em\lower.7ex\hbox{E}\kern-.125emX}}
\usepackage{balance}
\hypersetup{hypertex=true,
    colorlinks=true,
    linkcolor=ccr,
    anchorcolor=ccr,
    citecolor=ccr}
\begin{document}
\title{Evolutionary Trigger Detection and Lightweight Model Repair Based Backdoor Defense}

\author{Qi Zhou, Zipeng Ye, Yubo Tang, Wenjian Luo, \IEEEmembership{Senior Member, IEEE}, \\ Yuhui Shi, \IEEEmembership{Fellow, IEEE} and Yan Jia
\thanks{This study is supported by the National Key R\&D Program of China (Grant No. 2022YFB3102100), University Stability Support Program of Shenzhen (Grant No. GXWD20231130113127003), Shenzhen Fundamental Research Program (Grant No. JCYJ20220818102414030), the Major Key Project of PCL (Grant No. PCL2022A03), Shenzhen Science and Technology Program (Grant No. ZDSYS20210623091809029), Guangdong Provincial Key Laboratory of Novel Security Intelligence Technologies (Grant No. 2022B1212010005).
\textit{(Corresponding author: Wenjian Luo)}}
\thanks{ Q. Zhou, Z. Ye, Y. Tang, W. Luo and Y. Jia are with Guangdong Provincial Key Laboratory of Novel Security Intelligence Technologies, School of Computer Science and Technology, Harbin Institute of Technology, Shenzhen 518055, Guangdong, China. W. Luo and Y. Jia are with the Peng Cheng Laboratory, Shenzhen 518055, Guangdong, China.
(e-mail: {22s051036, 22b351009, 22s151061}@stu.hit.edu.cn, luowenjian@hit.edu.cn, jiayanjy@vip.sina.com)}
\thanks{Y. Shi is with School of Computer Science and Engineering, Southern University of Science and Technology, Shenzhen 518055, Guangdong, China.
(e-mail: shiyh@sustech.edu.cn)}
}

\markboth{}%
{How to Use the IEEEtran \LaTeX \ Templates}

\maketitle

\begin{abstract}
Deep Neural Networks (DNNs) have been widely used in many areas. However, DNN model is fragile to backdoor attacks. A backdoor in the DNN model can be activated by a poisoned input with trigger and leads to wrong prediction, which causes serious security issues in applications. It is challenging for current defenses to eliminate the backdoor effectively with limited computing resources, especially when the sizes and numbers of the triggers are variable as in the physical world. We propose an efficient backdoor defense based on evolutionary trigger detection and lightweight model repair. In the first phase of our method, CAM-focus Evolutionary Trigger Filter (CETF) is proposed for trigger detection. CETF is an effective sample-preprocessing based method with the evolutionary algorithm, and our experimental results show that CETF not only distinguishes the images with triggers accurately from the clean images, but also can be widely used in practice for its simplicity and stability in different backdoor attack situations. In the second phase of our method, we leverage several lightweight unlearning methods with the trigger detected by CETF for model repair, which also constructively demonstrate the underlying correlation of the backdoor with Batch Normalization layers. Source code will be published after accepted.
\end{abstract}


\begin{figure*}
\centering
\includegraphics[scale=0.4]{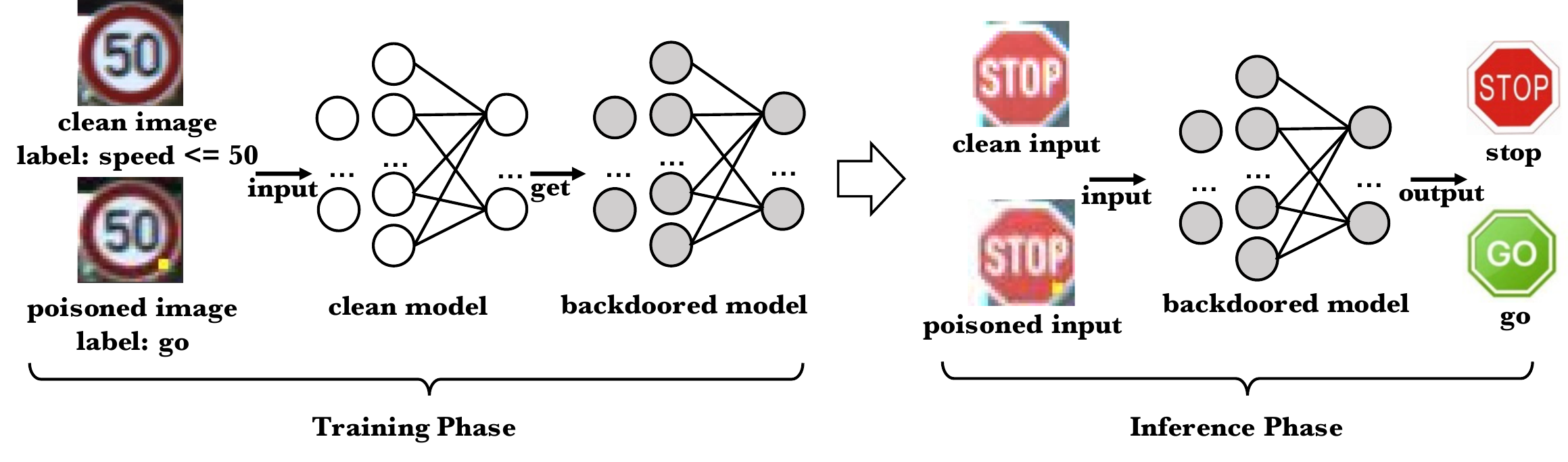}
\caption{The overview of the backdoor attack. Backdoor is inserted to the model by poisoned images with triggers (yellow patch in the figure) and target labels (label `go' in the figure) in the training phase. And in the inference phase, when an input is poisoned with the trigger, the model outputs the target label. The performance of classifying clean inputs without triggers is not influenced.}
\label{bd}
\end{figure*}

\section{Introduction}

\IEEEPARstart{D}{eep}  Neural Networks (DNNs) have been increasingly used in various areas such as traffic sign recognition in autonomous vehicles \cite{selfdriving}, medical image analysis in healthcare \cite{medical}, face recognition in security systems \cite{facerecog}  and risk assessment in finance \cite{finance}. 

However, DNNs are susceptible to malicious attacks. Backdoor attacks are a particularly insidious and powerful type of attacks that can cause DNNs to exhibit normal behaviors for clean samples while displaying abnormal behaviors when the samples are poisoned with triggers \cite{BadNets,Trojan}. Poisoned samples with triggers are always classified to the target label by a backdoored DNN model as shown in Fig. \ref{bd}. Backdoor attacks cause wide applications of DNNs in risks. For example, a backdoored model in an autonomous vehicle would misclassify the traffic signs with triggers, which can cause a serious security accident.
Despite the significant impacts, the backdoor is difficult to detect since abnormality only emerges when there are triggers in inputs and defenders usually have no prior knowledge of the triggers. 

Obviously, defense against the backdoor attacks is of great significance. Excellent defense methods are expected to eliminate the backdoor effect while maintaining the performance of predicting clean inputs. Existing defenses employ different ways to defend against backdoor attacks.
For example, many of them aim to eliminate the backdoors by fine-tuning or pruning the models \cite{finetuning,finepruning}. The classic Neural Cleanse technique modifies the model using trigger reversing and neuron pruning technologies, which are effective against backdoor attack but compromise classification accuracy \cite{NeuralCleanse}. 
There are also defenses based on sample-preprocessing that do not need to modify the structure of the parameters of the models. STRIP blends clean images linearly with the input, then determines whether the input is poisoned by observing the divergence of the predictions on blended images \cite{STRIP}. However, it is difficult to set a suitable divergence threshold which can always separate clean and poisoned inputs accurately. Moreover, Februus \cite{februus} neutralizes the backdoor effect by removing the most influential part depending on GradCAM \cite{gradcam} and restoring it by Generative Adversarial Network (GAN) \cite{GAN}. This defense processes all the inputs, whether they are poisoned or clean, so it inevitably affects the model performance on clean inputs since GAN cannot always repaint the removed part accurately. NEO locates the position of the trigger by repeatedly randomly cropping a part from the input and pasting it to other clean images to see if the number of changed predictions exceeds a threshold \cite{NEO}. However, such a random search based method is not precise and effective, and the unsuitable threshold also influences the model usability.

Besides, another limitation of existing methods is that they cannot handle triggers with different sizes and numbers, while these kinds of triggers are common in physical world.
Clearly, it is challenging to mitigate the backdoor effect while maintaining the model performance. 

To achieve these objectives simultaneously, we propose an efficient backdoor defense based on evolutionary trigger detection and lightweight model repair. The trigger detection is achieved by a sample-preprocessing based method, i.e., CAM-focus Evolutionary Trigger Filter (CETF).  In the inference phase of the model, CETF checks every image input of the model and filters the poisoned images with triggers. The triggers extracted by CETF are then used to repair the model by our unlearning methods. 

Specifically, in CETF, we first detect influential regions roughly by GradCAM \cite{gradcam} for the input.
Then, we design the objective function rationally and solve it based on an evolutionary algorithm \cite{evolutionary}, so as to accurately locate the most influential part of the input image with a size as small as possible.
After that, we paste the optimized region obtained by the evolutionary algorithm on a set of clean images and check whether their predictions change, thus determining whether there is a trigger in the input.
CETF is able to satisfy both the robustness and the usability requirements, and even robust with the triggers variable in sizes and numbers. 

After CETF, we unlearn the backdoored model by images with the extracted trigger for model repair. The naïve unlearning method is based on fine-tuning the model with the images pasted the reversed triggers \cite{NeuralCleanse}. Benefiting from the extracted triggers of CETF which are so accurate that approximate the triggers in the training phase, our model repair by unlearning is lightweight and practical. Also, the unlearning can be done in a very short time.

Furthermore, we find the existence of backdoor in the Batch Normalization (BN) layers. Therefore, based on our findings, we propose two more efficient methods, BN-unlearning and BN-cleaning, for model repair. Specifically, by modifying only the parameters of the BN layers, our proposed model-repair methods get excellent effect in a large number of experiments, which further demonstrates the validity of our findings.
Our viewpoint that the backdoor is injected by attackers to the model by setting the statistical information in the BN layers when training with triggers has great significance to the interpretability of backdoor.

Our contributions can be summarized as follows.

\begin{itemize}
    \item We firstly combine the evolutionary algorithm with backdoor defense. CETF is an effective way to search for the trigger in poisoned inputs accurately. Experimental results demonstrate the high robustness of CETF against the content, size and other attributes of triggers, thereby proving its universality and applicability in practice.
     \item The defense based on extracting triggers by CETF and repairing the model by unlearning shows a superior effect. We also propose two lightweight model repair methods for efficiency. Substantial experimental results show that our defense mitigates the backdoor effectively and efficiently without influencing the inference performance of the model.
     \item Furthermore, we explore the ``shortcut'' for trigger in the backdoored model. We propose that the backdoor is hidden in the Batch Normalization layers and a lot of experiments are conducted for verification. Based on the excellent effect of our backdoor defense, we think that our findings are inspiring to the future research about AI backdoors and model's interpretability.
\end{itemize}

\section{Related Work}
\subsection{Backdook Attacks}
There are many different kinds of backdoor attacks currently. BadNets is the representative of the patch-based visible backdoor attack \cite{BadNets}. The attacker poisons training images by pasting an arbitrary trigger on them and changing their labels to a target label. After training with poisoned images and clean images, the model behaves normally when the input is clean but outputs the target label when there is a trigger in the input in inference phase.

Different from BadNets whose trigger is selected arbitrarily, Trojan Attack generates the trigger by changing values of pixels in a trigger mask to achieve maximum values of some chosen neurons \cite{Trojan}. Trojan Attack assumes the attackers have no access to original training data, so it reverses training data and combines them with the trigger to retrain the model. After retraining, the trojan model could misbehave when an input is poisoned by the same trigger.

Both of BadNets and Trojan Attack poison images with patch-based visible triggers, and there are also some works about other visible triggers and invisible triggers which are more stealthy \cite{invisible,invisible2}.
The Input-Aware Attack generates triggers varying from image to image, so different inputs are poisoned with different triggers for stealthiness\cite{inputaware}.
Blended Injection is a backdoor attack blending the pixels of an input image and the trigger to generate a poisoned image \cite{blending}, and it balances inconspicuousness and backdoor effect by a blending weight.
One of the backdoor attack methods based on the invisible triggers is SIG\cite{sig}. SIG superimposes a backdoor signal such as horizontal sinusoidal signal to the images for poisoning, and it poisons images without modifying the labels to achievie great stealthiness. 
WaNet is another invisible backdoor attack \cite{wanet}. WaNet uses warping-based trigger, and the difference is unnoticeable so that the poisoned image is almost the same as the original image for human eyes.

However, these stealthy attacks, to a greater or lesser extent, are vulnerable in the physical world \cite{vunphy,physicalbackdoor}. The triggers in the poisoned images of the inference phase cannot be guaranteed to be the same as those in the training phase since the former are more fragile when captured by camera in real time. For example, supposing there is a printed traffic sign with a trigger on the side of the road and aiming to mislead the backdoored driving system maliciously,  the distance between the camera of the driving system and the traffic sign as well as the shooting angle could influence the picture, and the illumination can even destroy the stealthy triggers, so the stealthy trigger on the traffic sign is easy to lose effect.

In the physical world, patch-based triggers are more solid and more likely to bring risk, so patch-based triggers need more attention for a securer physical environment.
In our experiments, we use two classic patch-trigger based attacks, i.e., BadNets and Trojan Attack, to evaluate the performance of our defense. We also use the more stealthy non-patch based attacks for verification of our viewpoint about the relationship of the backdoor and the BN layers.

It is widely believed that the backdoor attack provides a shortcut for trigger in the model \cite{shortcut_poe}\cite{fewshot}\cite{honeypot}. Therefore, an image poisoned with  a trigger goes through the shortcut whatever the original content of the image is, thereby being classified as the target class. 

However, there has not been a consensus on the mechanism of backdoor in the model structure. Yang et al. have analyzed the mechanism of backdoor \cite{suppress}. They think the skip connections influence backdoor and they try to suppress the skip connections for defense. However, suppression has obviously negative effects on model's inference performance. Therefore, their experiments only demonstrate the importance of the skip connections for the model's usability, but cannot demonstrate directly that the backdoor is highly correlated to the skip connections. 

In this paper, we propose a viewpoint that the backdoor is hidden in the BN layers and demonstrate its validity by applying our defense on various backdoor attacks.

\subsection{Backdoor Defense}
Backdoor defense is a very significant research topic. We simply divide them into two kinds: one is repairing the model by adjusting the architecture or parameters, and the other is  preprocessing the samples before inputting them to the backdoored model.

Fine-pruning is a model-repair based defense combining pruning and fine-tuning, which detects the backdoor neurons through the activations of neurons in the final convolutional layer and prunes them \cite{finepruning}, and then fine-tunes the neurons through retraining with clean samples. The performance of classifying clean inputs is inevitably influenced after pruning neurons, because the final convolutional layer is important to the model usability, and in our view, the backdoor is hidden in the BN layers instead of the convolutional layers.

Neural Cleanse is the state-of-the-art backdoor defense \cite{NeuralCleanse} based on model repair. It believes that the backdoored model could misclassify images to the target label by adding certain perturbations, and misclassifying them to other labels needs more perturbations. Thus researchers calculate the average perturbation for every label. Then they determine the outlier perturbation as a reversed trigger and the corresponding label as the target label. After that, they repair the DNN model by neuron pruning or unlearning. However, neuron pruning affects the model's performance for clean inputs, and its fine-tuning needs high computation cost, which are also the drawbacks of many model-repair based defenses including Fine-pruning.

As a sample-preprocessing based defense, STRIP applies linear blend strategy to an input image and a set of clean images, and then determines whether the input is clean or poisoned through the diversity of blended images' predictions \cite{STRIP}. Specifically, the predictions of images blended with clean input tend to be more diverse, while those corresponding to poisoned input are more likely to be the target label. However, when the trigger is broken in the blend process, the divergence cannot be obvious enough to distinguish poisoned inputs completely. Unlike blending pixels linearly, our CETF substitutes the pixels on clean images by pixels of the trigger to ensure triggers' intactness.

Februus initially utilizes GradCAM to generate the salient map, and removes influential regions from the input based on the salient map \cite{februus}. Subsequently, it employs GAN to repaint the removed regions. Februus performs indiscriminate removing and repainting on every input, regardless of whether the data is clean or poisoned. Therefore, it is hard for Februus to maintain the accuracy on clean samples, as GAN cannot always precisely restore the removed regions. Furthermore, the threshold of the salient map directly determines the size of the region to be removed, but the defender typically does not know the size of the trigger, resulting in a lack of generalizability for a certain threshold.

NEO randomly searches for the backdoor trigger by blocking it to see whether it results in a transition of prediction \cite{NEO}. For confirmation, the region is extracted and placed on clean images to check if the number of their prediction transitions exceeds a threshold. However, the influential regions of clean inputs could also exceed the threshold because the threshold is calculated in the same way for randomly extracted regions from clean images, which leads to misdiagnosis for clean inputs. What's more, NEO is often ineffective and inefficient for its simple random search since triggers' positions and sizes are both unknown.

\section{Method}
\begin{figure*}
\centering
\includegraphics[scale=0.31]{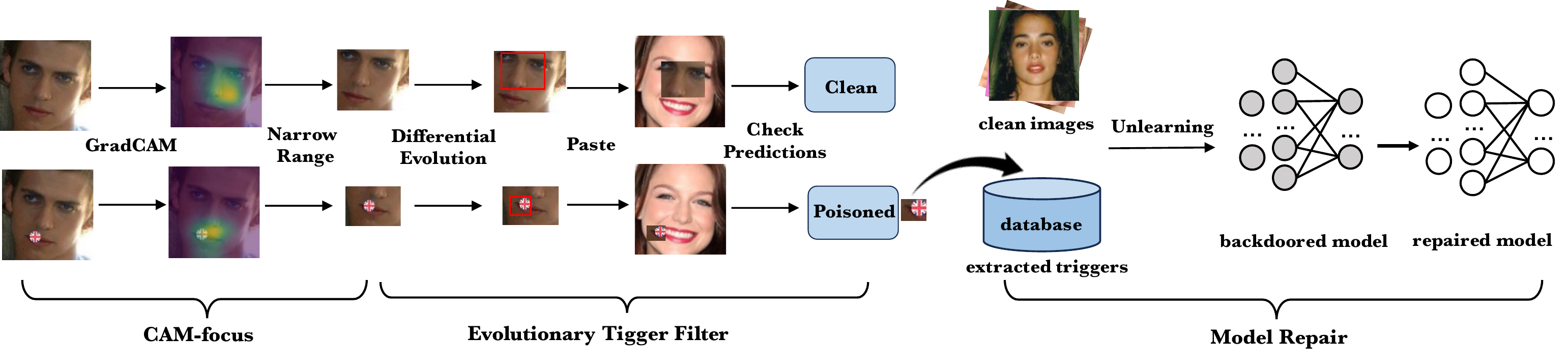}
\caption{Pipeline of our defense based on the CAM-focus evolutionary trigger filter and model repair. The first row of CETF is the process when the input is clean, and the second row is the process when the input is poisoned with a trigger (the British flag).
CETF is able to pinpoint the location of the triggers, thus helping to effectively differentiate poisoned samples from clean ones and use them to repair the backdoored model.}
\label{pipeline}
\end{figure*}
\subsection{Threat Model}
Backdoor attacks usually appear when a user does not have enough computation resources to train a DNN model so outsources it to a third party. As shown in Fig. \ref{bd}, if a malicious third party puts triggers in clean inputs to generate poisoned inputs and changes their labels to the target class in training phase, then the model could map the trigger as a feature to the target label. In the inference phase, when there is an input with the same trigger, the model will misclassify it to the target class. But when the input is clean, the backdoored model outputs prediction normally. Formally, for ($x_i$, $y_i$) belong to a valid dataset, if $x_i$ is a clean input to the backdoored model $f_\theta$ ($\theta$ represents the parameters of the model),  then $f_\theta(x_i)=y_i$, which is the true label, but $f_\theta(x_{i_p})=y_t$ where $x_{i_p}$ is poisoned and $y_t$ is the target label.

\subsection{Overview}
The pipeline of our method is shown in Fig. \ref{pipeline}. Our backdoor defense includes CAM-focus evolutionary trigger filter and model repair. CETF first narrows down the searching range for the trigger in the image by GradCAM, which can generate a salient map to present the importance of every region in the image \cite{gradcam}. Then the prior region got from the salient map becomes the main search range by the evolutionary algorithm. After detecting and validating the trigger by the evolutionary optimization approach, unlearning methods are used with the extracted triggers to repair the backdoored model. The details of each phase are as follows.

\subsection{CAM-Focus}
In the CAM-focus phase, we apply the GradCAM, which is an algorithm utilizing gradients of predictions combined with interpolation to generate a salient map \cite{gradcam}, to present the important regions in the images.
For a clean input image, the region containing important classification features will have larger weights than other regions in the salient map.
As for a poisoned input, the region containing the trigger is expected to have the largest weights because the trigger is the key part leading to a controlled prediction.
Therefore, important regions could be extracted by a preset salient map threshold.
Obviously, the threshold has a significant impact on the size of the extracted influential part, and if the threshold is inappropriate, the extracted part is easy to be either too small or too big, which means being an incomplete trigger or containing a large clean part of the image, respectively.
Also, GradCAM is not completely reliable because it does not always capture the entire influential object with a fixed threshold \cite{camplus}. As a result, depending only on GradCAM to extract the entire trigger accurately does not always work well, which is also the limitation of Februus. 

Therefore, in our method, GradCAM is utilized just to save time by roughly finding the influential region that will be used for the accurate trigger search in the next step, which means CETF can achieve the defense without dependence on GradCAM. 
In our experiments, we choose influential region with weights larger than 0.7 in the salient map and dilate its minimum bounding rectangle as the final output of CAM-focus prior. The threshold 0.7 need not be accurate since only a rough region is needed. We choose 0.7 because it represents a relatively large significance between 0 and 1; it is not affected by the scenarios so that the applicability is guaranteed.
Usually, this prior region contains the most influential part, which means classification features for the clean input or the trigger for the poisoned input. 
In the next step, the trigger can be precisely located inside it. 

\subsection{Evolutionary Trigger Detecting}

We utilize Differential Evolution (DE) \cite{DE2,DE} as the framework of our evolutionary optimization approach.
DE is an optimization algorithm based on the difference of the population individuals and shows stunning performance in many practical applications \cite{DEuse,DEuse2}. In CETF, customized DE is used to search for the minimum influential part of input image within the prior region got from CAM-focus. Specifically, based on DE, the minimum influential part can be obtained by initialization, mutation, crossover and selection. The implementation details are shown as follows.

\subsubsection{Initialization}
We randomly initialize the population $P$ with 40 individuals, i.e, $P=\{u_i\}_{i=1}^{40}$, where $u_i \in \mathbb{R}^4$.
Specifically, $u_i$ is used to determine a rectangular region, and the four entries of each $u_i$ represent the pixel position of the upper left corner of the searched region, as well as the height and width, respectively.

\subsubsection{Mutation}
The mutated individual $v_i$  of $u_i$  is generated by a differential process:
\begin{equation}\label{mutation}
 v_i(t)=u_i(t)+\alpha(u_{i_1}(t)-u_{i_2}(t))
\end{equation}
where $t$ represents current generation number, $u_{i_1}$, $u_{i_2}$ are individuals chosen randomly from the population and their difference is used to represent the difference of the current population individuals, and $\alpha$ is the scaling factor used to control the magnification degree of DE (we set $\alpha=0.3$ here).

\subsubsection{Crossover}
After mutation, offspring $\hat{u}_i(t)$ is generated by the mutated individual $v_i (t)$ and its parent individual $u_i (t)$ as follows:
\begin{equation}\label{*}
\hat{u}^{j}_i(t)=
\begin{cases}
v^{j}_i(t), &\gamma<p\\
u^{j}_i(t), &\gamma>p
\end{cases}
\end{equation}
where $(\cdot)^j$ represents the $j^{th}$ entry of vector $(\cdot)$, $\gamma$ is sampled from the uniform distribution $U(0,1)$, and the crossover coefficient $p$ is set to 0.5, which means parent individual and mutated individual are expected to have the same contribution to the offspring.
\subsubsection{Selection}
In the selection phase, we select individuals for new generation population $P^+$ by evaluating fitness values for each parent-offspring pair from current generation population and offspring generation, and the one which has the larger fitness function value $\text{fitness}(u)$ will be kept.
\begin{equation}\label{*}
    P^+=\bigcup_i\{u^{*}_i|u^{*}_i= \mathop{\arg \max}_{u\in \{\hat{u}_i,u_i\}}\text{fitness}(u)\}
\end{equation}
\subsubsection{Objective function}
We prefer the searched region to have backdoor effect and be as small as possible.
To achieve these objectives simultaneously, we design the fitness function as follows:
 \begin{equation}\label{*}
    \text{fitness}(u)=1000 \times \text{flag}(u)+1000 \times \text{flips}(u)-\frac{ \text{s}(u)^2}{\mathcal{S} }
\end{equation}                   
where $\text{flag}(u)$ is a binary value, and it returns 1 only if we substitute the region (determined by $u$) with average color of the user's input image and feed it into the model, causing the model's predicting label to change.
Otherwise $\text{flag}(u)$ returns 0.
And $\text{flips}(u)$ is calculated as follows:
we paste the region $u$ on auxiliary image set $\mathcal{D}_a$, and $\text{flips}(u)$ is the ratio of the predicted class in $\mathcal{D}_a$ that is the same as that of the user's input image.
For example, if there are $h$ auxiliary images in $\mathcal{D}_a$ (we set $h=10$ in our experiments) and $q$ of them change the predicted class to that of the user's input image after pasted the current region $u$, then $\text{flips}(u)$ is $q/h$. And $\text{s}(u)$ is the area of the searched region; $\mathcal{S}$ is the total area of the input image. 

Notably, $\text{flag}(\cdot)$ is designed to make sure the optimized region is important for prediction,
and $\text{flips}(\cdot)$ is designed to ensure the region can change the predictions of auxiliary images.
Both $\text{flag}(\cdot)$ and $\text{flips}(\cdot)$ correspond to the backdoor effect property of the trigger.
As for the area $\text{s}(\cdot)$, we control it to not only search for a block containing the trigger, but also avoid getting a very big block containing the whole influential part of a clean image. $\frac{ \text{s}(u)^2}{\mathcal{S} }$ is designed to ensure $\text{s}(\cdot)$ is controlled by considering the input image's area $\mathcal{S}$. The hyper-parameter 1000 in (4) is used to make the three terms in suitable order of magnitude to balance their influences.

We set the maximum number of evolutionary generations as 100 and Fig. \ref{deoptimization} shows that 100 generations are sufficient for convergence. When the best value of the fitness function has not improved for 10 consecutive generations, the trigger search ends in advance.  After 100 generations at most, for a poisoned input, we get the trigger, while for a clean input, we get a clean image feature, or nothing because the search ends with the value of the fitness function less than 0 which means DE cannot find an influential part from clean images. When DE cannot find an optimized region, we decide the current input as clean directly, so the image need not go to the next step, which saves time and computation resource smartly. 

As for other hyper-parameters in DE, detailed explanations and experiments can be found in the  \emph{Parameter Analysis} part of the experiments section.

DE process for trigger search is shown in details in Algorithm~\ref{trigger searching} and Algorithm~\ref{fitness}. 
Algorithm~\ref{trigger searching} shows the whole process including initialization, mutation, crossover and selection. 
 The $\textbf{RandInitial}$ function in line 3 of Algorithm~\ref{trigger searching} represents initializing population with 40 individuals randomly, while $\textbf{RandomSelect}(P)$ in line 9 represents selecting 2 individuals randomly from the current population for mutation in line 10. And $\textbf{Uniform}(0,1)$ in line 12 represents uniform distribution. The crossover procedure is shown from line 11 to line 15, and the selection is shown in line 27. The DE process ends if the best fitness value converges as shown in lines 20 to 24.
 
 \begin{algorithm}[H]
    \caption{Trigger Search.}
    \label{trigger searching}
    \textbf{Input}: a DNN model $f_\theta$, an input $x$\\
    \textbf{Parameter}: $\alpha$ = 0.3\\
    \textbf{Output}: $u_{best}$ 
    \begin{algorithmic}[1] %[1] enables line numbers
        \definecolor{light-gray}{gray}{0.45}  
        \STATE\textcolor{light-gray}{\emph{\# initialize P randomly}}
        \STATE\textcolor{light-gray}{\emph{\# each $u_i$ is a patch position in the input $x$}}
        \STATE$P \gets \textbf{RandInitial} \{u_i\}_{i=1}^{40}$       
        \STATE$u_{best} \gets \arg\max_{u \in P} \text{fitness}(u)$
        \STATE maxTrappedCount = 0
        \FOR {$t \gets 1, 2, ..., 100$}
        \FOR {each $u$ in $P$}
        \STATE\textcolor{light-gray}{\emph{\# select individuals from P randomly}}
        \STATE $u_{1},u_{2} \gets \textbf{RandomSelect}(P)$
        \STATE $ v \gets u+\alpha(u_{1}-u_{2})$
        \FOR{$j$ in range(4)}
        \STATE $\gamma \gets  \textbf{Uniform}(0,1)$
        \STATE\textcolor{light-gray}{\emph{\# $u=(u^1, u^2, u^3, u^4)$ decides a searched region}}
        \STATE $\hat{u}^j \gets u^j $ if $ \gamma > 0.5$ else  $\hat{u}^j \gets v^j$
        \ENDFOR
        \STATE\textcolor{light-gray}{\emph{\# the calculation of  fitness is shown in Algorithm 2}}
        \STATE $u$ = $\hat{u}$ if $\text{fitness}(u)<\text{fitness}({\hat{u})}$
        \ENDFOR
        \STATE\textcolor{light-gray}{\emph{\# break if convergent}}
        \IF{$\lvert \max_{u \in P} \text{fitness}(u)-\text{fitness}(u_{best)}\rvert < 1e-6$}
        \STATE maxTrappedCount $++$
        \IF{maxTrappedCount $>$10}
        \STATE break
        \ENDIF
        \ELSE
        \STATE maxTrappedCount = 0
        \STATE$u_{best} \gets \arg\max_{u \in P} \text{fitness}(u)$
        \ENDIF
        \ENDFOR  
        \STATE \textbf{return} $u_{best}$
    \end{algorithmic}
\end{algorithm}

The design of the fitness function is one of the keys of the trigger search by DE. Algorithm~\ref{fitness} shows the details of the calculation for fitness function clearly.
The $ \textbf{Mask}(x, u)$ function in line 2 of Algorithm~\ref{fitness} means masking $x$ on the position $u$, and $\textbf{Paste}(x_c, u)$ in line 5 is a function that pasting the suspected trigger patch on the position $u$ of $x_c$. 

\begin{algorithm}[h]
    \caption{fitness($u$).}
    \label{fitness}
    \textbf{Input}: a DNN model $f_\theta$, a clean validation dataset $V$, a patch position $u$  in an input $x$, area ${\mathcal{S}}$ of $x$ \\
    \textbf{Output}: $\text{fitness}(u)$
    \begin{algorithmic}[1] %[1] enables line numbers
        \definecolor{light-gray}{gray}{0.45}  
        \STATE\textcolor{light-gray}{\emph{\# mask $x$ on the position $u$}}
        \STATE $x^{-} = \textbf{Mask}(x, u)$
        \STATE $\text{flag} =  \mathbb I (f_\theta (x^-) \neq f_\theta(x) )$
         \STATE\textcolor{light-gray}{\emph{\# paste patch of $x$ to $x_c$ on the position $u$}}
        \STATE $V^+ = \{ {x_c}^+|{x_c}^+=\textbf{Paste}(x_c, u), x_c \in V\}$
        \STATE $\text{flips} =(\sum_{{x_c}^+\in V^+}  \mathbb I (f_\theta ({x_c}^+)== f_\theta(x)))/ \lvert V^+\rvert$ 
        
        \STATE $\text{fitness}(u) \triangleq 1000 \times \text{flag}+1000 \times \text{flips}-\frac{ \text{s}(u)^2}{\mathcal{S} } $ 
        \STATE\textcolor{light-gray}{\emph{\# s(u) is the area of u}}
        \STATE \textbf{return} $\text{fitness}(u)$
    \end{algorithmic}
\end{algorithm}

\subsection{Pasting and Filtering}
To determine whether the influential part is a trigger or feature of a clean input, we place it on a set of clean images on the same position as its original location on the input image as shown in Fig. \ref{pipeline}.
We find that a trigger is much stronger and more independent than clean image feature, which means a trigger can lead to the target label even though the trigger is incomplete or pasted on an unrelated image.
As a result, we record and check the predictions of images pasted by the optimized region, and if the majority of predictions changes to the prediction of the original input image, the optimized part is a trigger and the input is poisoned, otherwise clean.

\subsection{Model Repair}
\subsubsection{Naïve Unlearning}
A common model repair method based on the trigger is unlearning the backdoor by fine-tuning. The fine-tuning process is similar to the model training process.
Neural Cleanse unlearns the backdoor by fine-tuning the infected model with images patched by the reversed triggers, and the labels of these images for unlearning are set to the true labels instead of the target label. After unlearning,  the model restores the inference performance and could output correct labels for input images even when the real triggers are present in the inference phase. It can be explained that when fine-tuning with the images from different classes with triggers and their true labels, the model will not associate the trigger as a feature to the target class, so it achieves backdoor mitigation.

In our method, we use the trigger extracted by CETF, instead of the reversed trigger in Neural Cleanse, to repair the backdoored model. The extracted trigger is also pasted to a set of clean images with true labels, and then these images are used to fine-tune the model to unlearn the backdoor. Since the CETF is expected to provide the optimized region which is more similar to the original trigger than the reversed trigger in Neural Cleanse, our unlearning phase is easy to get superior model repair effect with less computation and data resources than Neural Cleanse.
\subsubsection{BN-Unlearning and BN-Cleaning}
We also propose two new novel unlearning methods based on modifying only the statistical information and parameters of the BN layers.

In almost all the state of the art computer vision tasks based on Convolutional Neural Network  \cite{resnet20,inception1,inception2,efficientnet}, the BN layer is almost an essential component since it can accelerate the convergence of networks and improve training speed and stability.
Batch Normalization can normalize the input along each feature dimension within a batch during training, and BN layer is typically placed before the activation function to ensure that the input data of each layer remains within a stable distribution \cite{BN}. 

In a BN layer, batch mean estimates the mean of each feature and batch variance estimates the variance of each feature in the current data batch, and they are used to normalize the input for each feature dimension.
Then the normalized features are scaled and shifted with the scaling parameter $\gamma$ and offset parameter $\beta$, thereby enhancing the model's expressive power and training effectiveness.
In the training phase, both of running mean $\mu$ and running variance $\sigma^2$ accumulate updates based on the statistics of the batch data, while the scaling parameter $\gamma$ and offset parameter $\beta$ are updated by back propagation typically with gradient descent. 
In the model inference phase, the input is not always a batch so it is necessary to utilize the accumulated running mean $\mu$ and running variance $\sigma^2$ updated during the training process to maintain consistent data distribution and normalization effects. 

There are researchers reversing clean images dependent on the BN layers in the backdoored model since they assert that only a small portion of the training data are poisoned so the information of the BN layers is not influenced \cite{fewshot}. However, there is not any experimental result to demonstrate this viewpoint in their paper. Besides, since there is large accuracy degradation in their experiments due to the difference between the reversed data and the clean data, we question their viewpoint that  the information of the BN layers is not influenced in backdoor training. On the contrary, in this paper, we do adequate experiments to demonstrate that the backdoor is closely related to the BN layers. In the experiments, the specific new unlearning methods are as follows.

\paragraph{BN-Unlearning} Given our belief that backdoor is hidden in the BN-layers, we propose to unlearn the backdoor by updating the parameters of the BN layers only, with parameters of other layers frozen. 
We believe that the parameters of the BN layers are shaped by the statistical information of the training data, which typically consists of a small proportion of poisoned images and a large proportion of clean images. Specifically, in the training phase, the mixture of the poisoned images and the clean images influences some specific parameters of the BN layers. After training, when inputting a poisoned image, the specific activations are greater than 0 after Batch Normalization and activate the model to output the target class, which is exactly the shortcut.
However, when inputting a clean image, the specific activations in the BN layers is less than 0, which cannot activate the model to output the target class. 

We get the poisoned images by inserting the extracted trigger from CETF to the local validation images, and then use these images with their true labels for unlearning by fine-tuning the model (here, only the parameters in the BN layers are tuned). After fine-tuning, the statistical information in the BN layers is obviously reset. Once the inputs with the trigger cannot get a large activation with the new parameters in the BN layers, the backdoor is unlearned successfully. 

\paragraph{BN-Cleaning} 
As the relative magnitudes of the activations in BN layers are mainly controlled by the running mean $\mu$ and running variance $\sigma^2$, we propose another unlearning method, BN-cleaning, which only modifies $\mu$ and $\sigma^2$, and the experiment of BN-cleaning can convincingly confirm the strong relationship between the backdoor and statistical information in the BN layers.

Different from the scaling parameter $\gamma$ and offset parameter $\beta$ that need to be updated by the gradient back propagation process, the running mean $\mu$ and the running variance $\sigma^2$ can be updated only by the images during the forward propagation process.
Therefore, in BN-cleaning, we only need to input the images pasted by the extracted trigger, and then the statistical information $\mu$ and $\sigma^2$ are updated automatically. 
The running mean $\mu$ and running variance $\sigma^2$ are updated as follows:
\begin{equation}\label{updatemu}
\mu_{new} =(1-\lambda) \cdot \mu_{old} + \lambda \cdot \mu_{batch}
\end{equation}
\begin{equation}\label{updatesigma}
\sigma^2_{new} =(1-\lambda) \cdot \sigma^2_{old} + \lambda \cdot \sigma^2_{batch}
\end{equation}
where $\lambda$ balances the old and new values;
$\mu_{new}$ and $\sigma^2_{new}$ are the updated running mean and updated running variance, respectively;
$\mu_{old}$ and $\sigma^2_{old}$ are the previous running mean and previous running variance, respectively;
$\mu_{batch}$ and $\sigma^2_{batch}$ are the sample mean and sample variance of the current batch, respectively.

It is obvious that the update of $\mu$ and $\sigma$ is independent of the loss function, so there is even no need to know the true labels of the images, which means BN-cleaning is applicable even we only have images without labels. Since we freeze all the parameters of other layers except for the BN layers and do not need back propagation, the computation cost is far lower than the naïve unlearning. 

\section{Experiments}
\subsection{Settings}

\subsubsection{Metrics}

We use two metrics, clean image accuracy and the attack success rate, to evaluate the maintenance of classification performance and the mitigation of the backdoor effect after defense, respectively. Model's classification accuracy Accu. and the attack success rate ASR are calculated as follows:
 \begin{equation}\label{*}
Accu.=\frac{1}{n} \sum_{i=1}^n \mathbb I {(f_\theta(x_i)=y_i)}
\end{equation}       
 \begin{equation}\label{*}
ASR=\frac{1}{m} \sum_{j=1}^m \mathbb{I} {(f_\theta(x_j)=y_t)}
\end{equation}  
where $f_\theta(x)$ is the model's prediction for image $x$, $n$ is the number of clean images, $m$ is the number of poisoned images with triggers, $y_i$ is the groundtruth label of the $i^{th}$ image and $y_t$ is the target label of the backdoor attack. And $\mathbb{I}{(\cdot)}$ is the indicator function.
Defense algorithms and backdoor repair methods are expected to keep Accu. high and reduce ASR as low as possible at the same time.

\subsubsection{Models and Datasets}
To evaluate the defense, we consider two widely used settings: face recognition and traffic sign classification. For face recognition, we choose FaceScrub \cite{FaceScrub} as the dataset, while for traffic sign classification, we choose GTSRB \cite{GTSRBdata,GTSRB} as the dataset.

We experiment with patch-based trigger attacks and non-patch based trigger attacks respectively, and for non-patch based attacks, we introduce the settings in the \emph{Additional Experiments for Verification} part. For patch-based attack, we use two benchmark backdoor attacks, BadNets \cite{BadNets} and Trojan Attack \cite{Trojan}. For BadNets, we use GTSRB to train ResNet20 \cite{resnet20} and FaceScrub to train MobileNet (a face classifier taking MobileNet as backbone which is widely used for face classification) \cite{mobilenets}. For Trojan Attack, we use VGG16, which is pretrained on 2,622,000 images with faces from 2,622 persons \cite{facerecog}, and then fine-tuned the last full connected layer of VGG16 model on clean images from FaceScrub.
VGG16 was introduced in 2014 \cite{vgg16} and was earlier than the work of Batch Normalization, so the original VGG16 model does not incorporate BN layers. Although VGG16 is not widely used in practice now and there are modified versions VGG16-BN,  we still decide to use VGG16 to achieve the attack performance similar to the original work proposing Trojan Attack \cite{Trojan}.

Our used GTSRB is a dataset consisting of 40,736 traffic signs of 43 kinds, and images are resized to $32 \times 32$.
Our used FaceScrub is a face dataset with 45,333 images of 526 persons.
For BadNets, we resize the images in FaceScrub to $160 \times 160$, while for Trojan Attack, we resize the face images to $224 \times 224$.
For each attack, we randomly choose 10\% of the training data to poison with the trigger and mix the poisoned images with the other 90\% images to train the models.

\subsubsection{Triggers and Defenses}
For Trojan Attack, the triggers are generated dependent on the models. For BadNets, the trigger of GTSRB is set as a yellow block with size $2 \times 2$ put on the bottom right corner fixedly, while the trigger of FaceScrub is a circular British flag logo with radius 20. Instead of putting the British flag triggers on the fixed positions, we put the triggers randomly so different images could have triggers on different positions, which can further demonstrate the robustness of our defense. The poisoned samples with triggers are shown in the Table \ref{settings}.

The accuracy for classifying clean images (Accu.) and the attack success rate (ASR) of the attacked models are shown in Table \ref{settings}.
From Table \ref{settings}, the original Accus. before defense are 99.97\%, 95.52\% and 93.30\% for ResNet20, VGG16 and MobileNet, respectively. And the corresponding ASRs without defense are 99.18\%, 100\% and 99.39\%, respectively. 

For comparison, we select 4 state-of-the-art defense algorithms: Neural Cleanse, STRIP, NEO, Februus.
\begin{table}[h]
    \centering
    \renewcommand\arraystretch{1.3}
    \caption{Experiment settings for backdoor attacks. We choose 3 models as well as the corresponding datasets and triggers. The Accus. of models and ASRs before defense are presented.}
    \label{settings}
    \begin{tabular}{cccc} % 控制表格的格式
    \toprule[1pt]
    \multicolumn{1}{c}{\multirow{2}[0]{*}{\ }} & \multicolumn{3}{c}{\textit{Backbone of Model}} \\
                
        & \multicolumn{1}{c}{ResNet20} & \multicolumn{1}{c}{VGG16} & \multicolumn{1}{c}{MobileNet} \\
    \midrule[0.3pt] 
    \textbf{Dataset}& GTSRB & FaceScrub & FaceScrub \\
    \textbf{Accu.} & 99.97\%&  95.52\% &  93.30\% \\
    \midrule[0.3pt]
    \textbf{Poisoned}& 
    \begin{minipage}[m]{0.2\columnwidth}
        \centering
        {\includegraphics[width=0.6\textwidth]{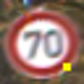}}
    \end{minipage} 
    &
    \begin{minipage}[m]{0.2\columnwidth}
        \centering
        {\includegraphics[width=0.6\textwidth]{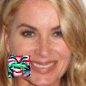}}
    \end{minipage} & 
    \begin{minipage}[m]{0.2\columnwidth}
        \centering
        {\includegraphics[width=0.6\textwidth]{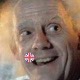}}
    \end{minipage} \\
    \textbf{Attack} & BadNets& Trojan Attack & BadNets\\ 
    \textbf{ASR} &  99.18\% & 100\% &   99.39\% \\
    \bottomrule[1pt]
    \end{tabular}
\end{table}

\subsubsection{Settings for Robustness}
Furthermore, we randomly change the number and size of the triggers to demonstrate the robustness of our CETF on the MobileNet model.
Specifically, to evaluate the robustness to the number of the triggers, we design a multi-trigger attack which poisons data by pasting 2 or 3 triggers randomly on the inputs from FaceScrub and evaluate the defense performance. As for the robustness to the size of the triggers, we train a new backdoored MobileNet model by setting every trigger's width and height randomly (in the interval between 10 and 50 for FaceScrub), which is called random-size-trigger attack. We select 1,000 samples from the test set of FaceScrub as the test set for the robustness experiment. In robustness experiment, we choose poisoned inputs which can attack DNN models successfully to evaluate defense algorithms, and the proportion of inputs which still attack successfully after defense is calculated as ASR.

\subsubsection{Implementation Details for Repair}
Once the trigger is extracted by CETF, it is added to the trigger database used to repair the backdoored model. For Facescrub and GTSRB, we choose 2 images for each class randomly from the local validation dataset and paste the extracted trigger on the images to construct a new dataset for repair.
After that, we fine-tune the backdoored model with the dataset for several epochs. 

During naïve unlearning, the parameters of all the model layers are updated by back propagation like the training phase, while BN-unlearning only updates the parameters of the BN layers and freezes other layers. For BN-cleaning, the operation is really easy that only updating the $\mu$ and $\sigma^2$ of the BN layers in the forward propagation is needed. Since there is no BN layer in VGG16, BN-unlearning and BN-cleaning are not conducted on the VGG16 models.

\subsubsection{Additional Experiments for Verification}
\begin{figure}
\centering
\includegraphics[scale=0.3]{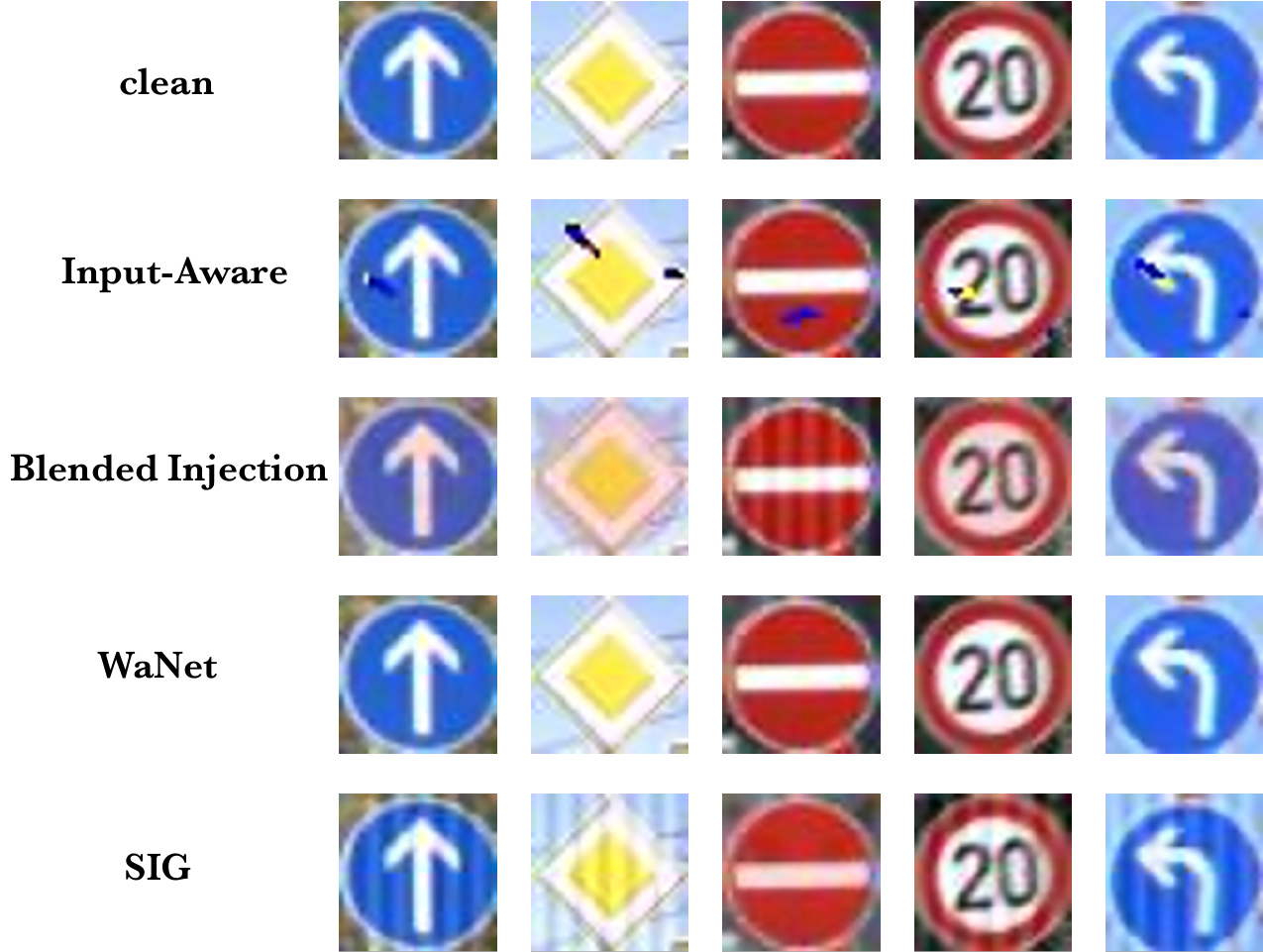}
\caption{Clean samples and the corresponding poisoned samples of Input-Aware Attack, Blended Injection, WaNet and SIG.}
\label{invisible}
\end{figure}
To demonstrate the close relationship of the backdoor and the BN layers, we apply unlearning, BN-unlearning and BN-cleaning to different kinds of backdoor attack: Input-Aware Attack \cite{inputaware}, Blended Injection \cite{blending}, WaNet \cite{wanet} and SIG \cite{sig}. Different from the BadNets and Trojan Attack, these attacks are non-patch based backdoor attacks. The Input-Aware Attack generates different triggers for different images, and Blended Injection, WaNet and SIG poison images by blending pixels with triggers, superimposing backdoor signals without modifying labels and applying warping-based triggers, respectively. The poisoned samples are shown in Fig. \ref{invisible}.

We poison the GTSRB and use the poisoned dataset to train the ResNet20 for the backdoored models of the above attacks respectively.
Then we fine-tune the backdoor by our unlearning methods to explore the inherent association of the backdoor and the BN layers. As for the number of samples for each class used for the fine-tuning, we experiment with different numbers from 1 to 4 to evaluate the relationship between the number of images and the model repair effect. 
Our experiments are implemented with NVIDIA TITAN RTX 3090 GPUs.

\subsection{Results}
In this part, we show our experimental results to demonstrate the performance of CETF and the model repair. For the performance of CETF, we focus on not only the defense effect against the backdoor attacks but also the DE process and the robustness. For the performance of model repair, we focus on not only the backdoor mitigation effect but also the relationship of the BN layers and the backdoors.
\subsubsection{Performance of CETF}
\begin{figure*}
\centering
\includegraphics[width=17cm]{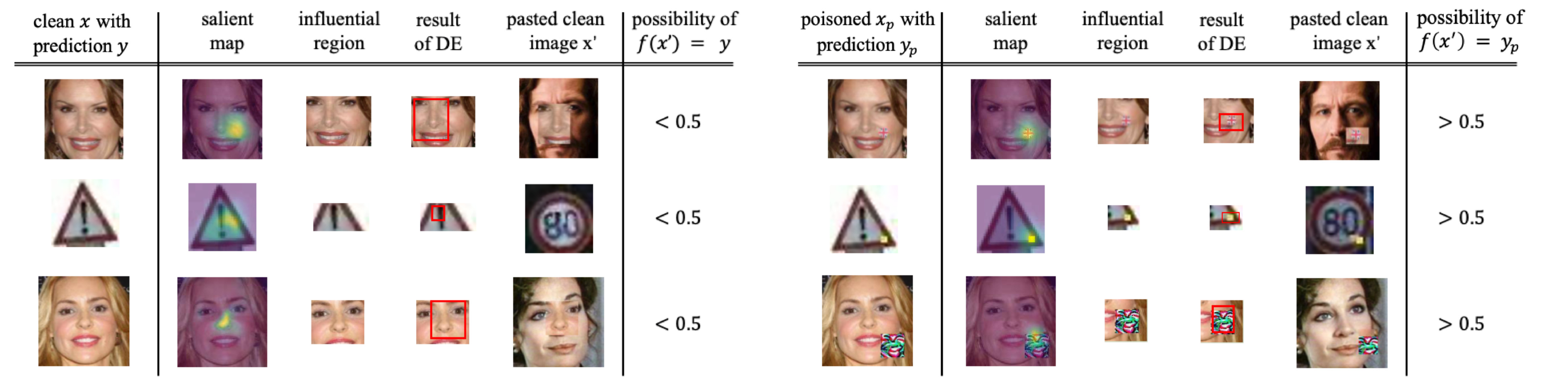}
\caption{The outputs of each step of CETF. The left shows the results of each step in the process when the inputs are clean, and the right is corresponding to the poisoned inputs.
Obviously, CETF can effectively distinguish between the clean inputs and the poisoned inputs, both for different data types and for different trigger types.}
\label{fig3}
\vspace{-3mm}
\end{figure*}

\paragraph{ASR and Accu. of CETF}
The visualization results of each step of CETF are shown in Fig. \ref{fig3}.
The Accus. and ASRs of the models after applying different defense methods are shown in Tables \ref{Accu} and \ref{ASR} respectively.

\begin{table}[h]
    \centering
    \renewcommand\arraystretch{1.2}
    \caption{Accus. of the backdoored models after defense. }  %表标题
    \label{Accu} 
    \begin{tabular}{l|rrrrr} % 控制表格的格式
    \toprule[1pt]
    \multicolumn{1}{l}{\multirow{2}[0]{*}{\textbf{Method}}} \vrule & \multicolumn{3}{c}{\textit{Backbone of Model}} \\
                
        & \multicolumn{1}{c}{ResNet20} & \multicolumn{1}{c}{VGG16} & \multicolumn{1}{c}{MobileNet} \\
    \midrule[0.3pt]
    \text{None} &99.97\%&95.52\%&93.30\%\\
    \midrule[0.3pt]
    \text{Neural Cleanse}&99.74\%&95.54\%&92.04\%\\
    \text{STRIP} &77.66\%&95.45\%&89.10\% \\ 
    \text{Februus}&77.02\%&89.59\%&75.56\%\\ 
    \text{NEO}&73.69\%&81.68\%&72.11\%\\ 
    \midrule[0.3pt]
     \text{CETF-only}&\textbf{99.97\%}&94.77\%&93.10\% \\ 
    \text{CETF+naïve unlearning} &\textbf{99.97\%}&\textbf{95.59}\%&\textbf{96.22\%} \\ 
    
    \bottomrule[1pt]
    \end{tabular}
\end{table}

\begin{table}[h]
    \centering
    \renewcommand\arraystretch{1.2}
    \caption{ASRs of backdoor attacks after defense.}  %表标题
    \label{ASR} 
    \begin{tabular}{l|ccccc} % 控制表格的格式
    \toprule[1pt]
    \multicolumn{1}{l}{\multirow{2}[0]{*}{\textbf{Method}}} \vrule & \multicolumn{3}{c}{\text{\textit{Backbone of Model}}} \\
                
        & \multicolumn{1}{c}{ResNet20} & \multicolumn{1}{c}{VGG16} & \multicolumn{1}{c}{MobileNet} \\
    \midrule[0.3pt]
    \text{None} &99.18\%&100\%&99.39\%\\
    \midrule[0.3pt]
    \text{Neural Cleanse} &\textbf{0}&7.63\%&6.49\%\\
    \text{STRIP} &55.04\%&0.19\%&15.55\% \\ 
    \text{Februus}&0.49\%&92.83\%&3.48\%\\ 
    \text{NEO}&12.93\%&13.96\%&7.21\%\\ 
    \midrule[0.3pt]
    \text{CETF-only}&\textbf{0}&\textbf{0}&0.72\%\\ 
    \text{CETF+naïve unlearning}  &\textbf{0}&\textbf{0}&\textbf{0}\\ 
    \bottomrule[1pt]
    \end{tabular}
\end{table}

There are two defenses with CETF in the tables, including CETF-only and CETF+naïve unlearning.
From Table \ref{Accu}, Neural Cleanse and the defenses with CETF can keep the model's usability in the experiments.
After defense by Neural Cleanse, the Accus. of the models have little decrease compared with those of the original backdoored models without defense. 
CETF-only also has a good maintenance of the Accus. for clean inputs.
The Accus. of the VGG16 and MobileNet models after the CETF+naïve unlearning even increase, which may be due to that our unlearning eliminates the backdoor in the models, thereby restoring the model performance to the clean models.

As for the reduction of ASR, the CETF+naïve unlearning is also obviously the best.
The high Accus. and low ASRs of CETF-only demonstrate the superior performance of CETF to distinguish poisoned images and clean images.
For ResNet20, Neural Cleanse and CETF perform best as  ASRs are 0. But for VGG16 and MobileNet, the ASR of Neural Cleanse are 7.63\% and 6.49\%, respectively, which are obviously worse than the effect of the two defenses with CETF.

Note that in the original experiments of Neural Cleanse \cite{NeuralCleanse},  the researchers create a new training dataset using 10\% samples of the original training data in which 20\% are added the reversed trigger with original labels, and then use the dataset to unlearn the backdoored model for just 1 epoch. The dataset used by them contains 2,622,000 images and the  ASR after unlearning is 3.70\%. As our dataset is much smaller,  we unlearn the model for 20 epochs for Neural Cleanse. It indicates that Neural Cleanse makes great demands on both the data and computation resources, which is not efficient. But our CETF and unlearning methods only need 2 images for each class, and unlearning for 3 epochs is adequate to get the superior effect, so the contribution lies in not only the better backdoor repair performance but also its better usability when the defenders have limited data and computation resources. 

Februus is even more unstable than Neural Cleanse because the ASR in VGG16 is as high as 92.83\%, which is caused by the drawback of GradCAM. Since the Trojan Square is relatively large, when the GradCAM cannot detect the entire influential region for Februus to remove, the remaining part will still work as trigger to attack the model. Similarly, STRIP reduces the ASR to 0.19\% in VGG16, but in ResNet20, the ASR is 55.04\% which is due to the trigger's vulnerability activated by the blending process. 
As for NEO, random search and unreasonable threshold lead to its weak performance.

\begin{figure}[h]
    \centering
    \includegraphics[scale=0.4]{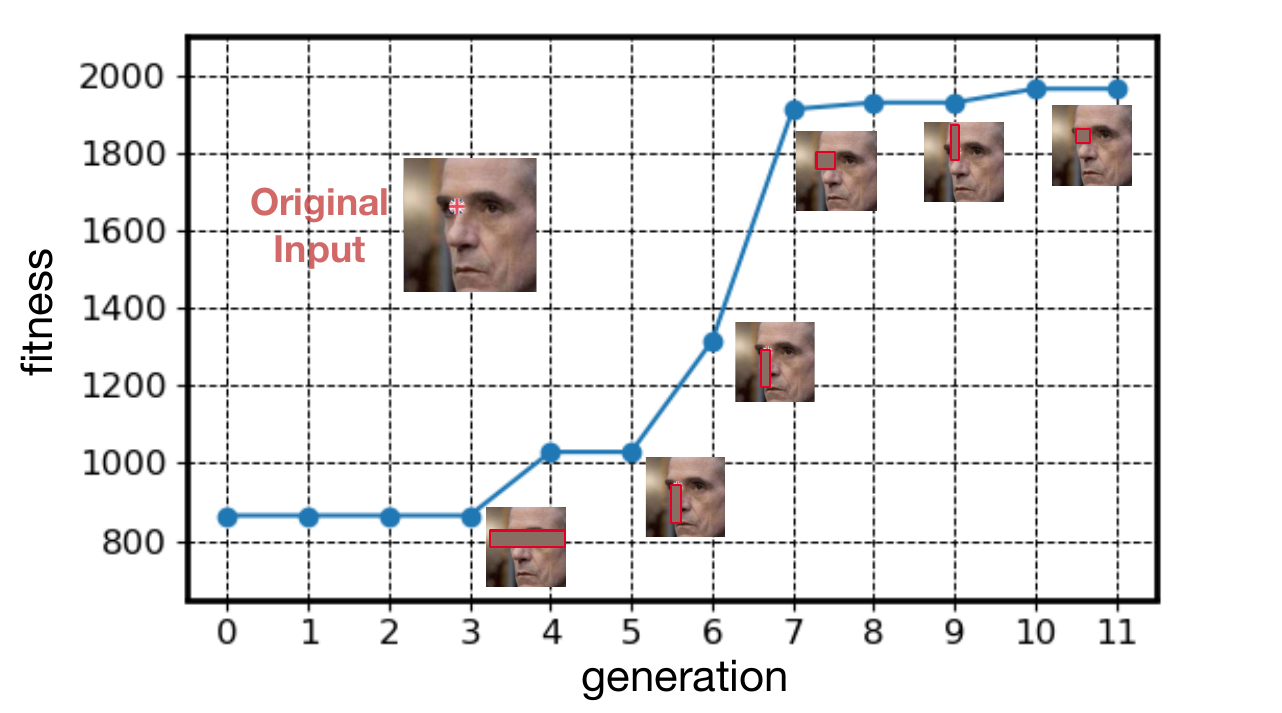}
    \caption{Optimization process of DE. DE optimizes the region generation by generation to get higher fitness values. A higher fitness value means a more accurate region containing trigger.}
    \label{deoptimization}
\end{figure}

Different from the instability of the compared defenses, defenses with CETF reduce the ASRs sharply. The effective mitigation of backdoor and good maintenance for model's performance show the excellent effect of CETF.

\paragraph{DE Process Analysis}
DE outputs the optimal individual with the largest fitness value as the final result, which represents the searched region that is most likely a trigger. We show the DE optimized process clearly in Fig. \ref{deoptimization}. The optimization  finishes after only 11 generations. From Fig. \ref{deoptimization}, as DE optimizes the region generation by generation, its position and size are more and more accurate to cover the trigger, which can be quantized by the higher and higher fitness value. The results show the excellent performance of DE. 

\begin{figure}[h]
\centering
\includegraphics[width=8cm]{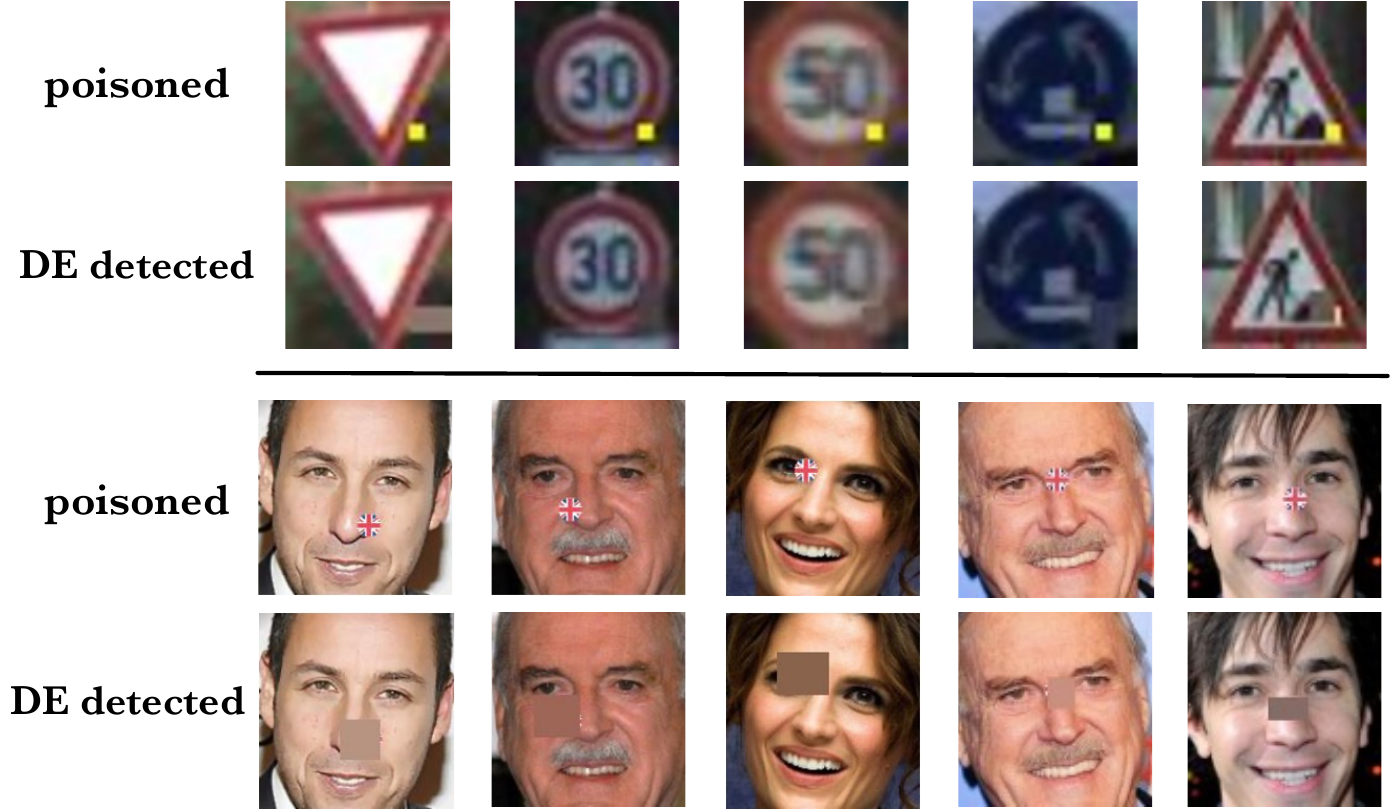}
\caption{Results of DE. The first row and the third row are poisoned images from GTSRB and FaceScurb, respectively; the second row and the fourth row are the corresponding results of DE, in which DE finds the triggers and blocks them with average color.}
\label{deresult}
\end{figure}

There are optimization results of DE for poisoned inputs shown in Fig. \ref{deresult}.
Triggers can always be found and blocked accurately by the evolutionary process, thus demonstrating the effectiveness of CETF.  

\begin{figure*}[t]
\centering
\includegraphics[scale=0.41]{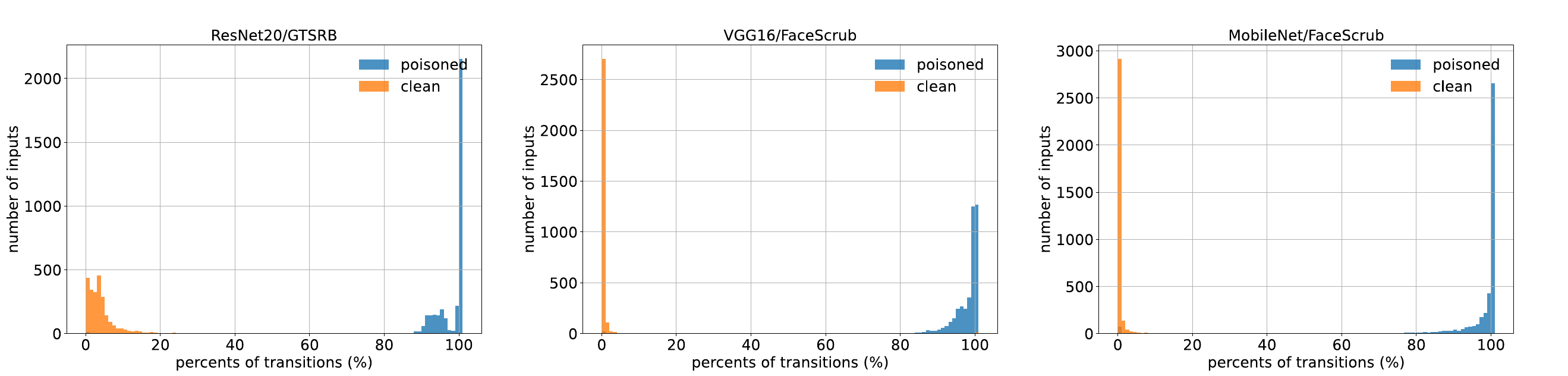}
\caption{The distribution of prediction transitions corresponding to clean and poisoned inputs of 3 models.
Clearly, for all experimental scenarios, there is always a large gap, which can effectively distinguish between clean and poisoned inputs.}
\label{threshold}
\end{figure*}

\paragraph{Parameter Analysis}
We explore the influence of DE's parameters on the performance of CETF and explain how we choose the values. We experiment with the number of individuals ranging from 10 to 50  and  $\alpha$ in Equation \ref{mutation} taking 0.1, 0.3, 0.5 0.7 and 0.9, respectively, with other parameters fixed. 

\begin{table}[h]
    \centering
    \renewcommand\arraystretch{1.5}
    \caption{ASRs after CETF's defense with different values for the number of individuals.}  %表标题
    \label{indi} 
    \begin{tabular}{c|rrrrrrr} % 控制表格的格式
    \toprule[1pt]
    \multicolumn{1}{c}{\multirow{2}[0]{*}{\textbf{Models}}} \vrule & \multicolumn{5}{c}{\text{\textit{Individuals}}} \\
                
        & \multicolumn{1}{c}{10} & \multicolumn{1}{c}{20} & \multicolumn{1}{c}{30} & \multicolumn{1}{c}{40} & \multicolumn{1}{c}{50}\\
    \midrule[0.3pt]
    \text{ResNet20} &15.49\%&3.21\%&0.38\%&0&0\\ 
    \text{VGG16} &7.26\%&0.68\%&0.02\%&0&0.02\%\\ 
    \text{MobileNet}&10.95\%&1.87\%&0.70\%&0.72\%&0.70\%\\ 
    \bottomrule[1pt]
    \end{tabular}
\end{table}

\begin{table}[h]
    \centering
    \renewcommand\arraystretch{1.5}
    \caption{ASRs after CETF's defense with different values for the scaling factor $\alpha$.}  %表标题
    \label{a} 
    \begin{tabular}{c|rrrrrrr} % 控制表格的格式
    \toprule[1pt]
    \multicolumn{1}{c}{\multirow{2}[0]{*}{\textbf{Models}}} \vrule & \multicolumn{5}{c}{\text{\textit{Scaling Factor $\alpha$}}} \\
                
        & \multicolumn{1}{c}{0.1} & \multicolumn{1}{c}{0.3} & \multicolumn{1}{c}{0.5} & \multicolumn{1}{c}{0.7} & \multicolumn{1}{c}{0.9}\\
    \midrule[0.3pt]
    \text{ResNet20} &1.30\%&0&0&0.38\%&2.40\%\\ 
    \text{VGG16} &0.05\%&0&0.05\%&0&1.23\%\\ 
    \text{MobileNet}&0.75\%&0.72\%&0.77\%&0.79\%&1.24\%\\ 
    \bottomrule[1pt]
    \end{tabular}
\end{table}

Table \ref{indi} shows how the number of individuals in a population affects the ASRs after CETF's defense. The ASRs of different models decrease consistently as the number of the individuals increases. It can be explained that more individuals could search for the triggers more sufficiently, so more poisoned inputs could be detected. When the number of individuals reaches 40, the trigger searching performance tends to be stable. The performance corresponding to 50 individuals is slightly better than that of 40 individuals for the MobileNet model, which could be attributed to the randomness in DE, and less individuals means less computation cost, so we choose 40 as the final decision for the number of individuals. 
As for the scaling factor $\alpha$ used to control the magnification degree of differential evolution in mutation as Equation \ref{mutation}, Table \ref{a} shows that 0.3 is the best choice. Since ASRs are also low with $\alpha$ taking the values round 0.3, it demonstrates our CETF's robustness to the parameter values.

\paragraph{Threshold Selection}
In the pasting and filtering phase, we stamp the influential part output by DE to a set of clean images and analyze their predictions.
If the ratio of predictions changing to the input's original prediction exceeds a threshold, then we determine the current input as poisoned.
Obviously, the threshold directly affects the performance of our defense, and we set the threshold to 50\% in our experiments. 

To demonstrate the validity and universality of our chosen threshold, we have counted the prediction changing ability of clean and poisoned samples, and the experimental results are shown in Fig. \ref{threshold}.
From Fig. \ref{threshold}, there is an obvious divergence between the prediction transitions corresponding to clean inputs and poisoned inputs.
And clearly, it is very easy for our threshold to distinguish poisoned inputs from clean inputs.

\begin{figure}[h]
    \centering
    \includegraphics[scale=0.34]{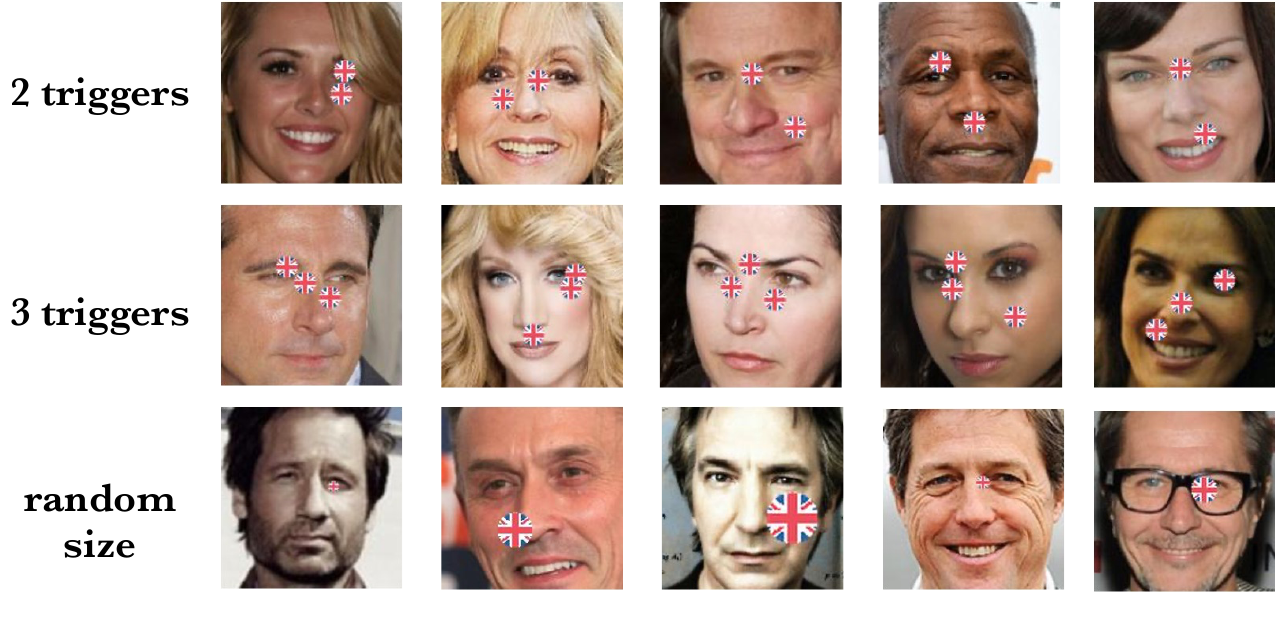}
    \caption{The poisoned samples with 2 triggers, 3 triggers and random-size triggers. These kinds of poisoned images are used to evaluate the robustness of the defense methods.}
    \label{robust}
\end{figure}
\subsubsection{Robustness Analysis}

\begin{table}[h]
    \centering
    \renewcommand\arraystretch{1.2}
    \caption{ASRs of backdoor attack variants after defense, which demonstrate the robustness of each defense to the number and size of triggers.}  %表标题
    \label{robustness}
    \begin{tabular}{l|ccccc} % 控制表格的格式
    \toprule[1pt]
    \multicolumn{1}{c}{\multirow{2}[0]{*}{\textbf{Method}}} \vrule & \multicolumn{3}{c}{\text{\textit{Attack Types}}} \\
                
        & \multicolumn{1}{r}{2 triggers} & \multicolumn{1}{r}{3 triggers} & \multicolumn{1}{r}{random size} \\
    \midrule[0.3pt]
    \text{Neural Cleanse} &20.98\%&41.17\%&12.75\%\\
    \text{STRIP} &0.90\%&0.30\%&15.66\% \\ 
    \text{Februus}&28.50\%&50.90\%&20.80\%\\ 
    \text{NEO}& 43.80\%&81.50\%&9.67\%\\ 
    \midrule[0.3pt]
    \text{CETF} \textbf{(Ours)}&\textbf{0}&\textbf{0}&\textbf{0.38\%}\\ 
    \bottomrule[1pt]
    \end{tabular}
\end{table}

In this part, we take a further step towards the multi-trigger attack as well as the random-size-trigger attack, and demonstrate that our defense method is highly robust to these backdoor attacks that are closer to the real scenarios.

\paragraph{Multi-Trigger Attack}
For the multi-trigger attack, we randomly paste 2 or 3 triggers to the validation data (from FaceScrub) as shown in Fig. \ref{robust} and feed them to the backdoored MobileNet.
From Table \ref{robustness}, the ASRs are 0 for poisoned inputs with 2 and 3 triggers after our CETF defense, which demonstrates that CETF is really robust to the multi-trigger attacks. More triggers make the trigger search in our CETF more efficient.
As for other methods, it is more difficult for Neural Cleanse to generate a reversed trigger to simulate all the trigger variants for unlearning thus it is likely to lose effectiveness. STRIP has a better defense performance than the single-trigger attack, since multi-trigger input can be distinguished more easily by STRIP due to the higher prediction possibility of multi-trigger input.
Februus is also limited since GradCAM cannot always highlight all the triggers simultaneously so there still remains triggers after Februus's removing operation.
NEO shows the worst performance, because when the multiple triggers distribute not closely, it is hard for NEO to find a size fixed block that covers all triggers, where the prediction will change if this block is removed.

\paragraph{Random-Size-Trigger Attack}
For the attack with random-size triggers, we poison data with triggers of random sizes and train a new backdoored MobileNet model with the data to evaluate the defenses. There are some samples poisoned with random-size triggers shown in Fig. \ref{robust}.
From Table \ref{robustness}, CETF presents the best performance.
Neural Cleanse's performance can still be explained by its lack of stability for trigger's variants.
STRIP nearly keeps its performance as that in backdoor attack of single fixed-size trigger.
Februus is likely to be out of action when the trigger is too large to be totally contained by GadCAM.
Compared to the other three methods, NEO shows a better performance, but it is still far inferior to our defense method.

\begin{figure*}[h]
    \centering
    \includegraphics[scale=0.28]{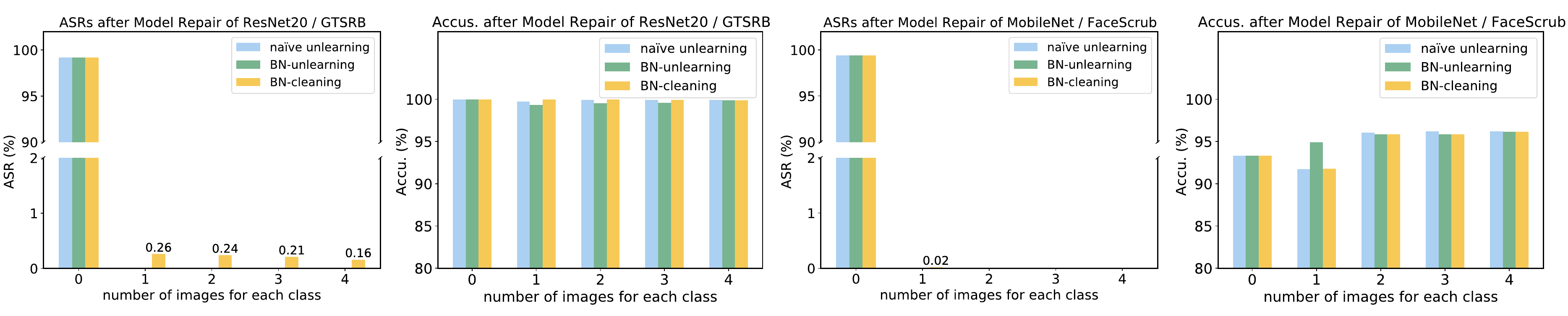}
    \caption{The ASRs and Accus. after repair by naïve unlearning, BN-unlearning and BN-cleaning on the backdoored ResNet20 and MobileNet models. The horizontal axis represents the number of images from each class used for unlearning. The image number of each class 0 on the horizontal axis is corresponding to the model's performance before repair.}
    \label{miti3}
\end{figure*}

\subsubsection{Performance of Repair}

In this part, we show the model repair results of the naïve unlearning, BN-unlearning and BN-cleaning.

From Fig. \ref{miti3}, these three repair methods get excellent effect  for each model with the extracted trigger from CETF. BN-unlearning can directly decrease ASRs of attacks to 0 and keep the models' classification accuracy by fine-tuning with only 1 image for each class, which demonstrates a striking applicability with limited resources. BN-cleaning cannot always decrease the ASRs to 0, but the ASRs after BN-cleaning with 1 image for each class, i.e., 0.26\% and 0.02\%, are also lower than many defense methods as shown in Table \ref{ASR}. 
The Accus. even become higher after our model repair methods compared to the Accus. before defense, and this may benefit from the outstanding backdoor-mitigation effect.
Considering these methods can all get stable results with only 2 images for each class, they are absolutely effective and practical repair methods, which also demonstrate the precise extraction for the triggers by CETF. Besides, the success of BN-unlearning and BN-cleaning also demonstrates the validity of our viewpoint about the relationship of backdoor and BN layers.

To further verify the viewpoint, we enrich the variety of backdoor attacks. Experiments on models attacked by Input-Aware Attack, Blended Injection, WaNet and SIG are added. Specifically, for each attack, we generate the corresponding poisoned samples for each class, and repair the backdoored model by the three repair methods. The repair effect is a powerful proof of our viewpoint.

\begin{table}[h]
    \centering
    \renewcommand\arraystretch{1.3}
    \caption{Accus. before and after repairing the backdoored model attacked by the various attacks.}
    \label{unlearningBA}
    \begin{tabular}{c|cccc} % 控制表格的格式
    \toprule[1pt]
    \multicolumn{1}{c}{\multirow{2}[0]{*}{\textbf{Repair}}} \vrule & \multicolumn{4}{c}{\textit{Backdoor Attack}}\\
                
        & \multicolumn{1}{c}{Input-Aware} & \multicolumn{1}{c}{Blended} & \multicolumn{1}{c}{WaNet}& \multicolumn{1}{c}{SIG}   \\
    \midrule[0.3pt] 
    before repair &99.94\%& 99.33\% & 99.67\%& 99.94\% \\
    \midrule[0.3pt] 
    naïve unlearning & 99.30\% & 96.21\%&  99.32\%& 99.56\% \\
    BN-unlearning & 99.87\%& 99.77\%& 99.87\%& 99.97\% \\
    BN-cleanning & 99.90\%&  93.01\%& 99.79\%& 99.97\% \\  
    \bottomrule[1pt]
    \end{tabular}
\end{table}

\begin{table}[h]
    \centering
    \renewcommand\arraystretch{1.3}
    \caption{ASRs  before and after reepairing the backdoored model attacked by the various attacks.}
    \label{unlearningasr}
    \begin{tabular}{c|cccc} % 控制表格的格式
    \toprule[1pt]
    \multicolumn{1}{c}{\multirow{2}[0]{*}{\textbf{Repair}}} \vrule & \multicolumn{4}{c}{\textit{Backdoor Attack}}\\
                
        & \multicolumn{1}{c}{Input-Aware} & \multicolumn{1}{c}{Blended} & \multicolumn{1}{c}{WaNet}& \multicolumn{1}{c}{SIG}   \\
    \midrule[0.3pt] 
    before repair & 97.98\%& 99.87\% & 99.67\%& 99.83\% \\
    \midrule[0.3pt] 
    naïve unlearning & 0 & 0 &  0 & 0 \\
    BN-unlearning & 0&  0 & 0 & 0 \\
    BN-cleaning & 0 & 1.1\%& 2.5\%& 0 \\
    \bottomrule[1pt]
    \end{tabular}
\end{table}

\begin{table}[h]
    \centering
    \renewcommand\arraystretch{1.2}
    \caption{Time cost of the 3 repair methods for each epoch.} 
    \label{time}
    \begin{tabular}{c|ccc} % 控制表格的格式
    \toprule[1pt]
    \multicolumn{1}{c}{\multirow{2}[0]{*}{\textbf{Model}}} \vrule & \multicolumn{3}{c}{\text{\textit{Repair Methods}}} \\
                
        & \multicolumn{1}{r}{naïve unlearning} & \multicolumn{1}{r}{BN-unlearning} & \multicolumn{1}{r}{BN-cleaning} \\
    \midrule[0.3pt]
    \text{ResNet20} &0.186s&0.154s&0.051s\\
    \text{MobileNet} &2.513s&1.860s&1.771s \\ 
    \bottomrule[1pt]
    \end{tabular}
\end{table}

From Table \ref{unlearningBA} and \ref{unlearningasr}, BN-unlearning presents near perfect effect for maintenance of model performance and mitigation of backdoor effect. From Table \ref{unlearningBA}, naïve unlearning and BN-cleaning also do well in performance maintenance, since there is only a little but not obvious decrease in Accus. From Table \ref{unlearningasr}, for every backdoor attack, the ASRs decrease directly to 0 after naïve unlearning and BN-unlearning. The ASR of the model attacked by Blended Injection and WaNet after BN-cleaning cannot decrease to 0, which can be explained that the BN-cleaning mainly updates the running mean $\mu$ and running variance $\sigma^2$, while the other 2 parameters $\gamma$ and $\beta$ are sometimes have some difference from those of the clean model and need to be updated to get perfect effect. However, it can still verify our viewpoint solidly. The excellent backdoor mitigation effect and maintenance of the model's classification performance verify successfully that repairing the BN layers could easily destroy the backdoors.

In addition to be a proof for our viewpoint, experimental results also show that BN-unlearning and BN-cleaning could be suitable repair methods in a defense pipeline after detecting the poisoned images.  Table \ref{time} shows the time cost for the repair methods. These unlearning methods are completed in a short time with our extracted triggers. For ResNet20, it costs less than 1s to restore the backdoored model for 3 epochs, which is really efficient. Although the MobileNet is trained by the images of faces from 526 classes, time costs of the 3 unlearning methods are all less than 3s for each epoch. Besides, it is obvious that BN-unlearning and BN-cleaning is more time-saving than naïve unlearning, since there are far less parameters to update.

Experimental results of BN-unlearning and BN-cleaning not only demonstrate the existence of backdoor in BN layers, but also provide an effective and efficient way for researchers  to repair the model by only updating the parameters in the BN layers. We believe this work will bring new insights and offer endless possibilities for research of backdoor attack and defense in the future.

\section{Conclusion}
In this paper, we propose a backdoor defense based on evolutionary trigger detection and efficient model repair by unlearning, which is inspiring for research of backdoor defense.
In detail, the trigger detection method CETF redesigns the key optimization factors of Differential Evolution to search for the trigger in a poisoned input and pastes it to clean images for verification.
The accurate trigger extracted from the poisoned input, as well as the huge decrease of the attack success rate, highly demonstrates the effectiveness of CETF.
And the robustness testing experiments further show that CETF is robust and practical. Using the trigger extracted by CETF, the unlearning methods repair the backdoored models successfully. Besides, we explore the mechanism of backdoor and believe our proposed BN-unlearning and BN-cleaning could guide the understandings of backdoor attacks.

Since the BN unlearning and BN cleaning repair the backdoored model by modifying the distribution of the statistical information in the BN layers, we will try to explore if there is other data except for the images with triggers that can influence the distribution, which could make the repair more convenient. Also, we hope to analyze the specific influence of triggers on the BN layers in the training phase theoretically in the future work. Besides, though current computer vision models usually have BN layers, we still want to explore the backdoor in the model without BN layers for a deeper understanding of backdoor attacks.

\bibliographystyle{IEEEtran}
\bibliography{references}

%Biography text here without a photo.
%\begin{IEEEbiography}[{\includegraphics[width=1in,height=1.25in,clip,keepaspectratio]{fig1.png}}]{IEEE Publications Technology Team}
%In this paragraph you can place your educational, professional background and research and other interests.\end{IEEEbiography}

\end{document}